# Comparison between Decoherence time for a Two-state Spin ½ System with Its Corresponding Quantum Retrieval Period

Afshin Shafiee[†] and Arash Tirandaz[‡]

Research Group on Foundation of Quantum Theory and Information,
Department of Chemistry, Sharif University of Technology,
P.O.Box 11365-9516, Tehran, Iran

**Abstract**: We consider the evolution of a two-state quantum system (a spin ½ particle) in both the framework of standard quantum mechanics and under the decoherence regime. The former approach on this issue is the well-known quantum flipping process of a dichotomic system subjected to a time-dependent magnetic field. In the latter approach, utilizing the Spin-Boson model to describe the interaction of system with its environment, we derive the Born-Markov master equation to obtain the decoherence time. It is possible to show that under some legitimate conditions, one may find a potential conflict between the predictions of decoherence theory and the result observed in a typical quantum flipping experiment.

## 1. Introduction

Undoubtedly, quantum mechanics is one of the most powerful theories in the history of physics. It has been widely used in practical applications. If one considers the consistency between the predictions based upon a theory and the experimental evidences as a criterion of the legitimacy of that theory, quantum mechanics has been extremely successful since no conflict (inconsistency) has been found to the present time. However, it is still a controversial issue that whether quantum mechanics is able to represent a perspicuous and unequivocal explanation about how the universe functions. When we demand quantum mechanics to present lucid description of the microscopic world, several difficulties arise.

The mysterious nature of the wave function might be the source of some of the existing conceptual problems. According to the linear dynamics of quantum equations, the wave function may be represented by a linear combination of several quantum states. It infolds some kind of coherency between the opposite parts of the description in itself. This wondrous feature of quantum mechanics, known as quantum superposition, manifests itself in the measurement problem. In a measurement, the entangled state of the system-apparatus evolves according to the deterministic Schrodinger equation. The ultimate state is another superposition comprised of entangled states of system-apparatus, corresponding to a specific outcome of measurement. But this is in sharp contrast with our everyday experience as an observer. We never find our classical world in a superposition of possibilities. Therefore, the measurement problem in its essence, faces the problem of transition from the microscopic world, represented by delicate quantum superpositions, to the classical world usually described by robust pointer states of a measuring apparatus.

[†] Corresponding author: shafiee@sharif.edu

[‡] tirandaz@mehr.sharif.ir

There are several approaches for resolving the measurement problem. Standard quantum mechanics deals with this difficulty by postulating the collapse of the wave function as a stochastic, indeterministic process that occurs when the quantum system of interest interacts with the measuring apparatus which also behaves quantum mechanically. Collapse theories like GRW add stochastic terms into the Schrodinger equation to explain the non-unitary nature of the collapse of wave function[1]. In Bohmian mechanics the existence of a guidance wave determines the ultimate position of the particles during a measurement process in a casual manner[2]. On the other hand there are other approaches like many-world interpretation [3] and modal interpretation [4] in which no additional dynamics other than the unitary time evolution of quantum states is introduced to explain the measurement problem. Nevertheless, the philosophical aspects of these interpretations are more dominant than their physical trait.

Decoherence theory tries to explain the so-called transition from quantum realm to classical world by considering the openness of quantum systems. In the framework of decoherence, realistic quantum systems are not isolated from their environment. In spite of classical physics, the environment has a crucial effect on the evolution of the system. Here the role of the environment is twofold: First it determines the kind of interaction from which the suppression of interference terms between quantum states of the system results. It corresponds to the suppression of the off-diagonal elements of the reduced density matrix of the system under its interaction with surrounding. Second: during the ubiquitous interaction of the system-environment, some preferred states of the system, known as pointer states, are selected[5]. These are states which preserve their correlation under the decoherence process. They have the least entanglement with the environment and the quasiclassical properties of the system can be attributed to them. Along with selecting the pointer states, the preferred state problem which depends on our ability to redefine quantum states in bases incompatible with each other is also resolved[6].

Since the environment has a large number of degrees of freedom, the phase coherence of the system becomes inaccessible to the observer. In addition, it can be shown that only a small fragment of the environment is necessary for the suppression of the interference terms[7]. Hence, decoherence is an effective and very fast process. So, the suppression of quantumness of the system, can be considered irreversible for all practical purposes. This can be also illustrated by the master equation formalism which represents the non unitary nature of measurement problem in the framework of decoherence[8]. As a result, classical nature would be an emergent property of the continuous monitoring of the system interacting with its environment.

Moreover, the interaction of the system with its environment occurs independently of the presence of any observer. Since the information about pointer states of the system has deposited in the environment (according to the openness of the system), the observer can get aware of the state of the system without disturbing it. Thus, it is possible to attribute objective properties to the pointer states which are quantum states in essence[9].

However, there is no consensus among physicists that decoherence has solved the measurement problem. For example, Ghirradi and Bassi have shown that if one considers the linearity of the evolution of quantum systems and the orthogonality of pointer states of a macro-object, the superposition of pointer states of the apparatus is still allowed[10]. Pessoa mentions that a partial trace on environment degrees of freedom to obtain the reduced density matrix of the system is equivalent to statistical version of the collapse of wave function in standard quantum mechanics [11]. Also, Adler points out that when quantum mechanics is applied uniformly to a system which is in interaction with a measuring apparatus and its environment, some kind of contradiction will arise about the interpretation of the final states of the apparatus, after the measurement process is completed [12]. In addition, the transition from a pure quantum state to a mixed one does not solve the measurement problem by itself. It seems that another collapse mechanism or an additional interpretation is necessary to complete our description of the process. Hence, some pioneers of decoherence confess that decoherence program cannot purely explain quantum to classical transition just by applying the basic formalism of quantum mechanics. Rather, it might be finally interpreted in the framework of many world interpretation [13].

On the other side, there are situations where the behaviour of the system, even in the macroscopic realm, is due to the preservation of a coherent superposition, in contrast with what decohernce program demands. For example the stable structure of molecules demonstrates the preservation of their constituent hybrid forms. Here, interaction with the environment cannot destroy the molecular structure. Decoherence should explain how such stable patterns have been formed in nature and why they are robust against the effect of the environment. Another example is quantum erasing process where quantum behaviour of the system can be retrieved reversibly [14].

In this paper, we are going to present another physical model in which the so called slogan "decoherence is everywhere" appears to encounter some difficulties. In section 2, we describe the time evolution of a two-state quantum system under a time dependent sinusoidal oscillatory potential. The quantum state can be retrieved after a given time. Then in section 3, we consider the interaction of the system with its environment, according to the decoherence program. Our calculations show that the quantum state of the system should be decohered, before it can get back to its initial status. The results of our work are summed up in final section.

## 2. Quantum Approach

Let us consider a spin ½ system (e.g. an electron) prepared in one of its spin eigenstates along the z-direction (say $|+\rangle$) at $t = t_0$. Afterwards, the particle is subjected to a time-dependent magnetic field rotating in the xy-plane. The Hamiltonian of the system under the influence of an external field is given by:

$$H = \omega_0 \hat{S}_z + \gamma \cos(\omega t) \hat{S}_x + \gamma \sin(\omega t) \hat{S}_y \qquad (1)$$

where $\gamma = \dfrac{eB}{m_e c}$ ( $e > 0$ and B is a constant demonstrating the intensity of the magnetic field), $\hat{S}_x (\hat{S}_y)$ is the $x(y)$ *component* of the spin operator and $\omega_0 = \dfrac{eB_0}{m_e c}$ is the angular frequency characteristic of the two state system. Also, the frequency of rotation of the field is denoted by $\omega$. Hamiltonian (1) is a typical form to describe how a spin ½ system undergoes a succession of spin-flops $|+\rangle \rightleftharpoons |-\rangle \rightleftharpoons |+\rangle$. Quantum state of the system at $t > t_0$ can be represented as:

$$|\psi(t)\rangle = C_+(t)|+\rangle + C_-(t)|-\rangle \tag{2}$$

where $C_+(t)$ and $C_-(t)$ are time dependent probability coefficients. If the initial state of the system were $|+\rangle$ a simple calculation yields the probability of the system to be found in the state $|-\rangle$ as:

$$|C_-(t)|^2 = \frac{\gamma^2}{\gamma^2 + (\omega - \omega_0)^2} \sin^2 \left( \left( \frac{\gamma^2 + (\omega - \omega_0)^2}{4} \right)^{\frac{1}{2}} t \right) \tag{3}$$

It is Rabi's formula [15] used for evaluating the transition probability between two states $|+\rangle \rightleftharpoons |-\rangle$ known as the spin-flop phenomena. At the resonance condition when $\omega = \omega_0$ as apparent from the above relation, we would need a time duration denoted by $\tau = \frac{2\pi}{\gamma}$ to get back to the same initial state $|\psi(0)\rangle = |+\rangle$.

According to the standard formalism of the deoherence theory, our proposed system described by state (2) may be coupled to an environment, so that the interaction between the system and the environment completely dominates the intrinsic dynamics of the system and the environment, separately. This is the basic assumption of an environment-induced decoherence which will be explored in next section. Yet, if the basic states $|+\rangle$ and $|-\rangle$ of the system could become correlated with the corresponding states of the environment and if the overlap between these new states gradually would decohere as time passes (on the authority of what decoherence model demands), then will it be again possible for the system to return to its initial state after a definite time? In next section, we assume that our spin ½ particle is subjected to an environment including harmonic oscillators known as a reservoir of bosonic field modes. The decoherence time scale of such an interaction determines that whether the interaction between a single dichotomic system and a bosonic environment dominates the dynamics of the system alone and whether it makes the system be found in one of its eigenstates, even before the state (2) can get back to its initial state. If so, it seems mysterious (of course, if one admits the decoherence approach) that two-state systems like Masers and so forth can undergo a succession of spin-flops $|+\rangle \rightleftharpoons |-\rangle \rightleftharpoons |+\rangle$ without being decohered under the same conditions. It is worth noting that in decoherence approach it is assumed the realistic quantum systems are inevitably coupled with their environment. It is very hard, if not impossible, to engineer the environment in such a way that the quantum system could be considered completely isolated [16,17]. In addition, supposing decoherence as a process which could be controlled by the experimenter arbitrary, it is meaningless to attribute objective properties to pointer states and then to consider classicality as an emergent property resulted from the interaction of the realistic quantum systems with their surroundings.

# 3. Decoherence program

Now, let us assume that the two-state spin ½ system described in the previous section is not actually isolated, but is subjected to an environment of harmonic oscillators containing a large amount of bosonic field modes $[8, ch.5]$. This assumption, however, does not mean that the system is *prepared* to be in contact with a special kind of environment, controlled and regulated by us, in a way similar to other interaction-based descriptions in physics which depend on the role of an experimenter. Rather, it just means that we are modelling the environment to illustrate the real dynamics of the system different from what the standard formalism of quantum mechanics depicts. So, the existence of an environment which affects the system is always pre-supposed in a decoherence program, but what is different in various approaches is the kind of the environment considered for modelling.

In the spin-boson model discussed here, the total Hamiltonian is defined by:

$$\hat{H} = \hat{H}_S + \hat{H}_\varepsilon + \hat{H}_{\text{int}} \tag{4}$$

where $\hat{H}_S$ is the time-dependent part defined in (1). The asymmetry energy $\hbar\omega_0$ denotes the difference in energy of the system between the states $|+\rangle$ and $|-\rangle$. In (4), $\hat{H}_\varepsilon$ denotes the self-Hamiltonian of the environment which consists of harmonic oscillators. $\hat{H}_{\text{int}}$ describes the interaction of $\hat{S}_z$ coordinates of the system with a linear superposition of the position coordinates $\hat{q}_i$ of each harmonic oscillator in the environment, defined by:

$$\hat{H}_{\text{int}} = \hat{S}_z \otimes \sum_i c_i \hat{q}_i \tag{5}$$

where $c_i$ is a coupling constant. Here, we assume that the interaction of the system with the large environment is not strong. The weak-interaction assumption permits us to use the Born and Markov approximations. According to the former, the time development of the density matrix of the environment is negligible, compared to system's evolution. So, the main temporal changes in the density matrix $\hat{\rho}$ of the system-environment combination are due to the change in the density operator of the system $\hat{\rho}_s(t)$ and $\hat{\rho}_\varepsilon \cong \hat{\rho}_\varepsilon(0)$ for the environment. Consequently, one can suppose that:

$$\hat{\rho}(t) \cong \hat{\rho}_s(t) \otimes \hat{\rho}_\varepsilon \qquad \forall t \geq 0 \tag{6}$$

On the other hand, the Markov approximation means that during the time evolution of the density matrix of the system, quantum correlations between the parts of the environment are destroyed rapidly. In other words, the internal self-correlations of the environment are not significant and the environment forgets its history (i.e., it is being kept in equilibrium) during its interaction with the system. Accordingly, the Born-Markov master equation can be derived as:

$$\frac{d}{dt}\hat{\rho}_s(t) = -\frac{i}{\hbar}\left[\hat{H}_s, \hat{\rho}_s(t)\right] - \frac{1}{\hbar^2}\int\limits_0^\infty dt \left\{ \upsilon(t)\left[\hat{S}_z, \left[\hat{S}_z(-t), \hat{\rho}_s(t)\right]\right] - i\eta(t)\left[\hat{S}_z, \left\{\hat{S}_z(-t), \hat{\rho}_s(t)\right\}\right]\right\} \quad (7)$$

where $\hat{\rho}_s(t)$ is the reduced density operator of the system and $\hat{S}_z(-t)$ shows the time dependence (in reverse direction) of the z-spin operator in the interaction picture. The noise and the dissipation kernels $\upsilon(t)$ and $\eta(t)$ can be defined as $[8, ch.5]$:

$$\upsilon(t) = \int\limits_0^\infty d\omega_e J(\omega_e)\coth\left(\frac{\hbar\omega_e}{2k_B T}\right)\cos(\omega_e t) \quad (8)$$

and

$$\eta(t) = \int\limits_0^\infty d\omega_e J(\omega_e)\sin(\omega_e t) \quad (9)$$

In the thermal equilibrium, $T$ and $\omega_e$ are, respectively, the temperature and the oscillation frequency of the particles of the environment. The function $J(\omega_e)$ is called spectral density of the environment which in the spin-boson model generally supposed as:

$$J(\omega_e) = \omega_e e^{-\frac{\omega_e}{\lambda}} \quad (10)$$

where $\lambda$ is the cut off (maximum) frequency. The spectral density is nearly ohmic (i.e., $J(\omega_e) \propto \omega_e$) for $\lambda \gg \omega_e$. According to the Hamiltonian (1) the unitary operator has the following form:

$$\hat{U}(t) = \exp\left[(-i(\frac{\omega_0 t}{\hbar}\hat{S}_z + \frac{\gamma t}{2\hbar}e^{-i\omega t}\hat{S}_+ + \frac{\gamma t}{2\hbar}e^{i\omega t}\hat{S}_-))\right] \quad (11)$$

Using the relation $\hat{S}_z(t) = \hat{U}^\dagger \hat{S}_z \hat{U}$, one gets (see Appendix A):

$$\hat{U}^\dagger \hat{S}_z \hat{U} = a_1(t)\hat{S}_z + a_2(t)\hat{S}_x + a_3(t)\hat{S}_y + a_4(t)\hat{I} \quad (12)$$

where the coefficients $a_1$ to $a_4$ are time dependent function defined in relations (A-17) to (A-20) in Appendix A Inserting these relations into the master equation (7) and doing some algebra, we get:

$$\frac{d}{dt}\hat{\rho}_s(t) = -i\left[\hat{H}'_s, \hat{\rho}_s(t)\right] - \tilde{D}\left[\hat{\sigma}_z, \left[\hat{\sigma}_z, \hat{\rho}_s(t)\right]\right] + D_1 + D_2 \quad (13)$$

where,

$$H_s^{\cdot} = \frac{1}{2}\left(\hat{\sigma}_z\left(\omega_0 - \int\limits_0^\infty \upsilon(t)a_4(t)dt\right) + \hat{\sigma}_x\left(\gamma\cos\omega t - \int\limits_0^\infty \upsilon(t)a_3(-t)dt\right)\right)$$

$$+\hat{\sigma}_y\left(\gamma\sin\omega t + \int_0^\infty \upsilon(t)a_2(-t)dt\right)\right) \tag{14}$$

$$\tilde{D} = \frac{1}{4}\int_0^\infty a_1(t)\upsilon(t)dt \tag{15}$$

$$D_1 = \frac{1}{4}\int_0^\infty a_3(-t)\left(\zeta\hat{\sigma}_y\hat{\rho}_S(t)\hat{\sigma}_z + \zeta^*\hat{\sigma}_z\hat{\rho}_S(t)\hat{\sigma}_y\right)dt, \tag{16}$$

and

$$D_2 = \frac{1}{4}\int_0^\infty a_2(-t)\left(\zeta\hat{\sigma}_x\hat{\rho}_S(t)\hat{\sigma}_z + \zeta^*\hat{\sigma}_z\hat{\rho}_S(t)\hat{\sigma}_x\right)dt \tag{17}$$

According to the decoherence formalism of the spin-boson model, the first term in (13), shows the unitary evolution of the system. The second term, however, represents the decoherence effect which shows how the environment monitors the **z**-component of spin observable of the system. Hence, $\tilde{D}$ is the coefficient of the decoherence rate. The last terms $D_1$ and $D_2$ in the master equation (13) can be used for illustrating the decay process of the system which is not of our interest here. In relations (16) and (17), we define $\zeta = (\upsilon(t) + i\eta(t))$.

The off-diagonal matrix elements of $\hat{\rho}_S(t)$ decay at a rate given by $\tilde{D}$. This can be shown by focusing only on the decoherence term in (13):

$$\frac{d}{dt}\hat{\rho}_S(t) = -\tilde{D}\left[\hat{\sigma}_z,\left[\hat{\sigma}_z,\hat{\rho}_s(t)\right]\right] \tag{18}$$

The above relation can be expressed in $\hat{\sigma}_z$ basis. Then, using the relation $\langle n|\hat{\sigma}_z\hat{F} - \hat{F}\hat{\sigma}_z|m\rangle = (n-m)\langle n|\hat{F}|m\rangle$ for arbitrary $\hat{F}$ and $n,m = \pm 1$, one can show that:

$$\frac{d}{dt}\rho_S^{(01)}(t) = -\tilde{D}\rho_S^{(01)}(t) \tag{19}$$

and

$$\frac{d}{dt}\rho_S^{(10)}(t) = -\tilde{D}\rho_S^{(10)}(t) \tag{20}$$

where $\rho_S^{(01)}(t)$ and $\rho_S^{(10)}(t)$ are off-diagonal elements of $\hat{\rho}_S(t)$ in $\hat{\sigma}_z$ basis. Defining

$$\tau_d = \frac{1}{\tilde{D}} \tag{21}$$

as the time needed for the exponential decay of the off-diagonal elements $\rho_S^{(01)}(t)$ and $\rho_S^{(10)}(t)$, it would be possible to predict whether the state of the system (described in (2)) can really return to its initial state at a given time $\tau = \dfrac{2\pi}{\gamma}$, or its corresponding density matrix undergoes a decoherence process before the change of state. Obviously, if $\tau_d \leq \dfrac{2\pi}{\gamma}$, then no returning process occurs. In this case, the decoherence prediction falls in contradiction with what is actually observed in practice, known as the spin-flop process in electron spin resonance experiments.

To show the aforementioned contradiction, we should first evaluate $\tilde{D}$, using the relations (8), (A-17) and (15). Here we assume that $k_B T \gg \hbar\omega_e$, which means that the thermal energy $k_B T$ is much larger than the oscillating energy of each particle in the environment. At room temperature, the thermal energy $k_B T$ is nearly equal to $10^{-21} J$. This is sometime called the thermal equilibrium of the environment $[8, ch.5]$. Since in the interaction term in (5), the spin coordinate of the system is involved, it is legitimate to assume that $\hbar\omega_e \leq 10^{-24}$ with $\omega_e \leq 2\pi \times 10^{10} s^{-1}$ which lies within the range of the frequency of electron spin resonance spectrum. For higher frequencies the spin degree of freedom cannot be coupled to the position coordinates of particles in the environment. Because, if the energy of the system $E_s$ becomes negligible compared to the oscillating energy of the environment (i.e. , $E_\varepsilon^{OSE}$ corresponding to $\hat{H}_{\bar{\varepsilon}}$ in (4)), the latter energy dominates the former and the environment screens off the system, instead of being entangled with it for monitoring the spin property of the system. So, the magnitude of $E_s$ imposes limitations on the greatness of $E_\varepsilon^{OSE}$. In weak interactions, $E_\varepsilon^{OSE}$ could be even smaller than $E_s$, so that the term $\hbar\omega_e$ could be supposed to be much smaller than $k_B T$. As a result, in a spin-boson model, for each particle in ordinary temperature, one can assume that $coth\left(\dfrac{\hbar\omega_e}{2k_B T}\right) \approx \dfrac{2k_B T}{\hbar\omega_e}$. Thus, using the relations (8), (A-17) and (15), when we neglect the higher order of $\gamma^2$, the decoherence factor at the resonance situation $\omega = \omega_0$ could be written as (see Appendix B):

$$\tilde{D} = \frac{\pi k_B T \gamma^2}{8\hbar\omega_0^2}(1 - e^{-\frac{k}{\lambda}}) \tag{22}$$

Where $k = \left(\gamma^2 + \omega^2\right)^{1/2}$. Since, $\dfrac{\gamma}{\omega_0} = \dfrac{B}{B_0}$, the ratio of decoherence time $\tau_d$ to the retrieval period $\tau$ will be:

$$\frac{\tau_d}{\tau} = \frac{4\hbar\omega_0}{\pi^2 k_B T}\left(\frac{\left(\dfrac{B_0}{B}\right)}{1 - e^{\frac{-k}{\lambda}}}\right) \tag{23}$$

At room temperature and in the electron spin resonance (ESR) region (i.e., $\omega_0 \approx 10^{10} s^{-1}$ ), one gets $\frac{4\hbar\omega_0}{\pi^2 k_B T} \cong 10^{-4}$. Thus, on can conclude that:

$$\frac{\tau_d}{\tau} \cong 10^{-4}\left(\frac{\omega_0}{\gamma}\right)\left(1 + e^{-\frac{k}{\lambda}} + e^{-\frac{2k}{\lambda}} + ...\right) \tag{24}$$

where we have used the relation $\left(1 - e^{-\frac{k}{\lambda}}\right)^{-1} \cong 1 + e^{-\frac{k}{\lambda}} + e^{-\frac{2k}{\lambda}} + ...$ for $0 < e^{\frac{-k}{\lambda}} < 1$. The cut off frequency $\lambda$ should have the same order of magnitude $\omega_0$ and $k$ for the reasons mentioned above, so that the determinacy factor in (24) is $\left(\dfrac{\omega_0}{\gamma}\right)$. In essence, $\dfrac{k}{\lambda}$ can not goes towards zero $(\lambda \to 0)$, when the energy contribution of the environment is not negligible compared to the system. For having a weak external field in (1), it is sufficient to have $\frac{\gamma}{\omega_0} \cong 10^{-3}$. So, $\frac{\tau_d}{\tau} \cong 10^{-1}$ and before the system can turn back to its initial state, it should decohere which is not compatible with experimental evidences. For getting a consistent result, one must consider $\frac{\gamma}{\omega_0} \leq 10^{-5}$ In such circumstances, the external field is so weak that in (1) one expects $\hat{H}_S \cong \omega_0 \hat{S}_Z$ and no resonance effect will then be observed. The region of violation $\frac{\tau_d}{\tau} < 1$ with respect to $\frac{k}{\lambda}$ in three different values of $\frac{\omega_0}{\gamma}$ is sketched in Figure 1.

Under the assumption $coth\left(\dfrac{\hbar\omega_e}{2k_B T}\right) \approx \dfrac{2k_B T}{\hbar\omega_e}$, the Caldeira-Leggett master equation is derived [18]. It is widely used in modelling of decoherence and dissipation processes. Considering a specific model based on nondemolition monitoring of oscillators with their environment, G. J. Milburn and D. F. Walls have shown that the Caldeir-Legget master equation can be applied successfully to situations in which the high temperature assumption is not strictly fulfilled [19]. Nevertheless, we have examined the validity of our claim by considering the broader range of temperatures in Appendix B. Here, when we can replace $coth\left(\dfrac{\hbar\omega_e}{2k_B T}\right)$ with its Taylor's series. Then, decoherence factor at resonance condition will have the following form at resonance situation:

$$\tilde{D} = \frac{\pi k_B T \gamma^2}{8 \hbar \omega_0^2} \left( 1 - e^{\frac{-k}{\lambda}} G(T) \right) \tag{25}$$

where $G(T)$ is:

$$G(T) = 1 + \frac{2^2 B_1}{2!} \left( \frac{\hbar k^{'}}{2 k_B T} \right)^2 - \frac{2^4 B_3}{4!} \left( \frac{\hbar k^{'}}{2 k_B T} \right)^4 + \frac{2^6 B_5}{6!} \left( \frac{\hbar k^{'}}{2 k_B T} \right)^6 - \dots \tag{26}$$

In (26), $B_1, B_3, \dots$ are Bernoulli numbers. Since $\tilde{D} > 0$ , so , $G(T) < e^{\frac{k}{\lambda}}$ and for few order of magnitudes of $e^{\frac{k}{\lambda}}$, $G(T)$ has not a significant role.

Finally, We have derived the general form of the decoherence factor considering the higher orders of $\gamma^2$ for the whole range of temperature (see Appendix B). The resulted decoherence factor, however, does not significantly differ from (22) or (25) in its value.

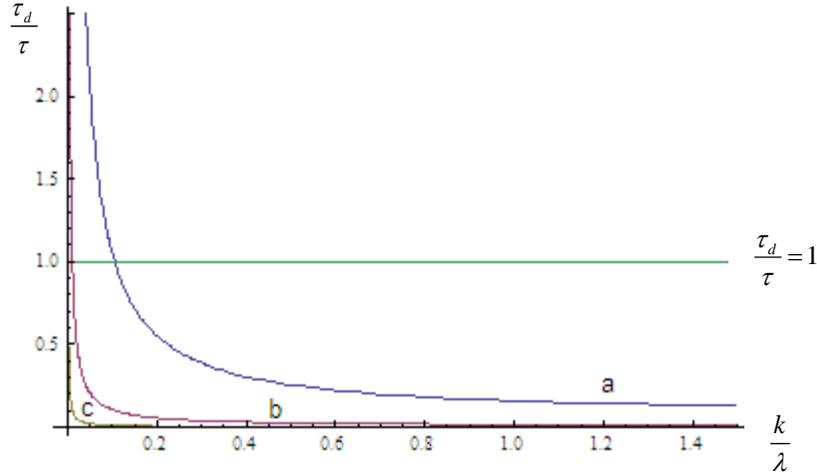

**Figure1**. $\frac{\tau_d}{\tau}$ is plotted against $\frac{k}{\lambda}$ for some different values of $\frac{\omega_0}{\gamma}$ equal to $a)10^3$ , $b)10^2$ and $c)10$ . In the region below the line $\frac{\tau_d}{\tau} = 1$ decoherence time is shorter than the retrieval period and the quantum system has not enough time to turn back to the initial state. As seen, even in very small portion of $\frac{B_0}{B} = \frac{\omega_0}{\gamma}$ it is still possible to find regions where the predictions of decoherence theory and standard quantum mechanics may display a lack of consistency in a typical electron spin resonance experiment.

## 4. Conclusion

At first sight, our conclusion may cause misconception. Here, we have explored a potential inconsistency in what a given decoherence model can predict for a well-known quantum process. Of course, standard quantum mechanics describe closed quantum systems while in decoherence models quantum systems should not be considered isolated from the environment. Why the predictions of the two theories should be compared with each other at all? The essence of any proper comparison between the predictions of quantum mechanics and its extended decoherence formalism lies on what is really observed in experience. Quantum mechanics neglects the role of the environment, but it does not mean that it denies the existence of the surrounding. Both theories presume the presence of the environment, but only in the decoherence program, the role and significance of environment is crucial. In this theory, if we accept the ubiquities interaction of the system with its environment, the density matrix of the system will be diagonlized due to the influence of its surrounding. As a result, the quantum coherence will be destroyed. The crucial point in this process is that, the observer has no determinant role in controlling the suppression of interference terms and the emergence of classicality. Since the transformation of pure states to mixed states is an irreversible process in the framework of decoherence, it should not be possible for the system to return to its initial state, when the decohering process is at work. However, this is exactly what is seen in the case of a two-state spin ½ system under a sinusoidal oscillatory potential. As usual, in real spin-flop phenomena, like ESR experiments, no body tries to isolate the quantum system from its surrounding by applying special techniques which is necessary to prevent the effect of the environment. On the other hand,  in our theoretical treatment of the above problem, the dominant role of the evolution of the system-environment comparing to the evolution of the system alone, is analyzed by the comparison between the retrieval time and the decoherence time. Our results show that under certain physical circumstances, if decoherence could work everywhere under the influence of physical parameters adjusted as expected (such as the cut off frequency of the environment), then no quantum state would be retrieved. This is in contrast to experimental evidences observed in the domain of ESR and NMR experiments.

## Acknowledgment

We would like to thank two anonymous referees of this journal for their valuable comments on the subject of the paper.

**Appendix A**

The main step in deriving the master equation is to calculate the evolution of $\hat{S}_Z$ operator in the interaction picture. According to the Hamiltonian (1) the unitary operator of the particle under the influence of the magnetic field is:

$$\hat{U}(t) = \exp\left[\left(-i\left(\frac{\omega_0 t}{\hbar}\hat{S}_z + \frac{\gamma t}{2\hbar}e^{-i\omega t}\hat{S}_+ + \frac{\gamma t}{2\hbar}e^{i\omega t}\hat{S}_-\right)\right)\right] \tag{A-1}$$

Using an auxiliary variable $\theta$ one could redefine the hermit conjugate of $\hat{U}^\dagger(t)$ as follows:

$$\hat{U}^\dagger(t) = \exp\left[\left(\left(\frac{i\theta}{\hbar}(\alpha_+\hat{S}_+ + \alpha_z\hat{S}_z + \alpha_-\hat{S}_-)\right)\right)\right]$$
$$= \exp\left(f_+(\theta)\frac{\hat{S}_+}{\hbar}\right)\exp\left(f_Z(\theta)\frac{\hat{S}_Z}{\hbar}\right)\exp\left(f_-(\theta)\frac{\hat{S}_-}{\hbar}\right) \tag{A-2}$$

where :

$$\alpha_+ = \frac{\gamma t}{2}e^{-i\omega t}, \alpha_- = \frac{\gamma t}{2}e^{i\omega t}, \alpha_z = \omega_0 t \tag{A-3}$$

and $f_-, f_+$ and $f_Z$ are some functions of $\theta$.

According to the disentangling theorem [21], for the particular case of $\theta = 1$, one gets the set of the following differential equations which connect the $\alpha_{\pm,z}$ coefficient to $f_{\pm,z}$ :

$$\dot{f}_+ - f_+\dot{f}_z - f_+^2\dot{f}_-\exp[-f_z] = i\alpha_+ \tag{A-4a}$$

$$\dot{f}_z + 2f_+\dot{f}_-\exp[-f_z] = i\alpha_z \tag{A-4b}$$

$$\dot{f}_-\exp[-f_z] = i\alpha_- \tag{A-4c}$$

where $\dot{f}$ means a differentiation with respect to $\theta$. Solving the above differential equations according to the constraint $f_i(0) = 0$ constraint, results the following relations for $f_{\pm,z}$:

$$f_+ = \frac{i\gamma e^{-i\omega t}}{\left(\gamma^2 + \omega_0^2\right)^{1/2}} \frac{\sin\left(\frac{t\left(\gamma^2 + \omega_0^2\right)^{1/2}}{2}\right)}{\cos\left(\frac{t\left(\gamma^2 + \omega_0^2\right)^{1/2}}{2}\right) - i\frac{\omega_0}{\left(\gamma^2 + \omega_0^2\right)^{1/2}}\sin\left(\frac{t\left(\gamma^2 + \omega_0^2\right)^{1/2}}{2}\right)} \tag{A-5}$$

$$f_- = \frac{i\gamma e^{i\omega t}}{\left(\gamma^2 + \omega_0^2\right)^{1/2}} \frac{\sin\left(\frac{t\left(\gamma^2 + \omega_0^2\right)^{1/2}}{2}\right)}{\cos\left(\frac{t\left(\gamma^2 + \omega_0^2\right)^{1/2}}{2}\right) - i\frac{\omega_0}{\left(\gamma^2 + \omega_0^2\right)^{1/2}}\sin\left(\frac{t\left(\gamma^2 + \omega_0^2\right)^{1/2}}{2}\right)} \tag{A-6}$$

$$f_z = -2\ln(\cos\left(\frac{t\left(\gamma^2 + \omega_0^2\right)^{1/2}}{2}\right) - i\frac{\omega_0}{\left(\gamma^2 + \omega_0^2\right)^{1/2}}\sin\left(\frac{t\left(\gamma^2 + \omega_0^2\right)^{1/2}}{2}\right)) \tag{A-7}$$

Then, one can write the $\hat{S}_z(t)$ in the interaction picture as:

$$\hat{S}_z(t) = \hat{U}^\dagger \hat{S}_z \hat{U} = e^{\frac{f_-}{\hbar}\hat{S}_+} e^{\frac{f_z}{\hbar}\hat{S}_z} e^{\frac{f_-}{\hbar}\hat{S}_-} \hat{S}_z e^{\frac{f_-^\dagger}{\hbar}\hat{S}_-} e^{\frac{f_z^\dagger}{\hbar}\hat{S}_z} e^{\frac{f_+^\dagger}{\hbar}\hat{S}_+} \tag{A-8}$$

The following steps leads to the final form of $\hat{S}_z(t)$ :

Step1:

$$\hat{A}_1 = e^{\frac{f_-}{\hbar}\hat{S}_-} \hat{S}_z e^{\frac{f_-^\dagger}{\hbar}\hat{S}_-} = \frac{\hbar}{2}\left(|+\rangle\langle+| + \left(1 - |f_-|^2\right)|-\rangle\langle-| + f_-^\dagger|+\rangle\langle-| + f_-|-\rangle\langle+|\right) \tag{A-9}$$

Step2:

$$\hat{A}_2 = e^{\frac{f_z}{\hbar}\hat{S}_z} \hat{A}_1 e^{\frac{f_z^\dagger}{\hbar}\hat{S}_z} \tag{A-10}$$

$$\hat{A}_2 = \frac{\hbar}{2}\left(\left(\exp\left(\frac{1}{2}(f_z + f_z^\dagger)\right)\right)|+\rangle\langle+| + (1 - |f_-|^2)\left(\exp\left(\frac{-1}{2}(f_z + f_z^\dagger)\right)\right)|-\rangle\langle-|\right.$$

$$+f_-^\dagger\left(\exp\left(\frac{1}{2}\left(f_z-f_z^\dagger\right)\right)\right)|+\rangle\langle-|+f_-\left(\exp\left(\frac{-1}{2}\left(f_z-f_z^\dagger\right)\right)\right)|-\rangle\langle+| \qquad \text{(A-11)}$$

Step3:

$$\hat{S}_Z(t)=e^{\frac{f_z}{\hbar}}\hat{A}_2 e^{\frac{f_z^\dagger}{\hbar}}=a_1(t)\hat{S}_z+a_2(t)\hat{S}_x+a_3(t)\hat{S}_y+a_4(t)\hat{I} \qquad \text{(A-12)}$$

Where $\hat{I}$ is the identity operator and $a_i(t)$ functions (i=1,2,3,4) are:

$$a_1(t)=\frac{1}{2}\left(|f_+|^2\left(1-|f_-|^2\right)\exp\left(-\frac{1}{2}\left(f_z+f_z^\dagger\right)\right)+f_+^\dagger f_-^\dagger\exp\left(\frac{1}{2}\left(f_z-f_z^\dagger\right)\right)\right.$$
$$\left.+\exp\left(\frac{1}{2}\left(f_z+f_z^\dagger\right)\right)+f_+f_-\exp\left(-\frac{1}{2}\left(f_z-f_z^\dagger\right)\right)-\left(\left(1-|f_-|^2\right)\exp\left(-\frac{1}{2}\left(f_z+f_z^\dagger\right)\right)\right)\right) \qquad \text{(A-13)}$$

$$a_2(t)=\frac{1}{2}\left(f_+^\dagger\left(1-|f_-|^2\right)\exp\left(-\frac{1}{2}\left(f_z+f_z^\dagger\right)\right)+f_+\left(1-|f_-|^2\right)\exp\left(\frac{-1}{2}\left(f_z+f_z^\dagger\right)\right)\right.$$
$$\left.+f_-^\dagger\exp\left(\frac{1}{2}\left(f_z-f_z^\dagger\right)\right)+f_-\exp\left(-\frac{1}{2}\left(f_z-f_z^\dagger\right)\right)\right) \qquad \text{(A-14)}$$

$$a_3(t)=\frac{i}{2}\left(f_+\left(1-|f_-|^2\right)\exp\left(-\frac{1}{2}\left(f_z+f_z^\dagger\right)\right)-f_+^\dagger\left(1-|f_-|^2\right)\exp\left(\frac{-1}{2}\left(f_z+f_z^\dagger\right)\right)\right.$$
$$\left.+f_-^\dagger\exp\left(\frac{1}{2}\left(f_z-f_z^\dagger\right)\right)-f_-\exp\left(-\frac{1}{2}\left(f_z-f_z^\dagger\right)\right)\right) \qquad \text{(A-15)}$$

$$a_4(t)=\frac{\hbar}{4}\left(|f_+|^2\left(1-|f_-|^2\right)\exp\left(-\frac{1}{2}\left(f_z+f_z^\dagger\right)\right)+f_+^\dagger f_-^\dagger\exp\left(\frac{1}{2}\left(f_z-f_z^\dagger\right)\right)\right.$$
$$\left.+\exp\left(\frac{1}{2}\left(f_z+f_z^\dagger\right)\right)+f_+f_-\exp\left(-\frac{1}{2}\left(f_z-f_z^\dagger\right)\right)+\left(\left(1-|f_-|^2\right)\exp\left(-\frac{1}{2}\left(f_z+f_z^\dagger\right)\right)\right)\right) \qquad \text{(A-16)}$$

Considering (A-5) to (A-7), each $a_i(t)$ coefficient could be rewritten as the following:

$$a_1(t) = \frac{1}{2}\left[ \frac{\frac{\gamma^2}{k^2}\sin^2\left(\frac{kt}{2}\right)}{\cos^2\left(\frac{kt}{2}\right)+\left(\frac{\omega_0}{k}\right)^2\sin^2\left(\frac{kt}{2}\right)}\left[\cos^2\left(\frac{kt}{2}\right)+\sin^2\left(\frac{kt}{2}\right)\left(\left(\frac{\omega_0}{k}\right)^2-\left(\frac{\gamma}{k}\right)^2\right)+\frac{\frac{k^2}{\gamma^2}}{\sin^2\left(\frac{kt}{2}\right)}-2\right]- \right.$$
$$\left. \cos^2\left(\frac{kt}{2}\right)+\sin^2\left(\frac{kt}{2}\right)\left(\left(\frac{\omega_0}{k}\right)^2-\left(\frac{\gamma}{k}\right)^2\right)\right] \qquad \text{(A-17)}$$

$$a_2(t) = \frac{\gamma}{k}\sin\frac{kt}{2}\left[\frac{\sin(\omega t)\cos\left(\frac{kt}{2}\right)-\frac{\omega_0}{k}\cos(\omega t)\sin\left(\frac{kt}{2}\right)}{\cos^2\left(\frac{kt}{2}\right)+\left(\frac{\omega_0}{k}\right)^2\sin^2\left(\frac{kt}{2}\right)}\right]\left[\cos^2\left(\frac{kt}{2}\right)+\sin^2\left(\frac{kt}{2}\right)\left(\left(\frac{\omega_0}{k}\right)^2-\left(\frac{\gamma}{k}\right)^2\right)-1\right] \qquad \text{(A-18)}$$

$$a_3(t) = \frac{\gamma}{k}\sin\frac{kt}{2}\left[\frac{\cos(\omega t)\cos\left(\frac{kt}{2}\right)+\frac{\omega_0}{k}\sin(\omega t)\sin\left(\frac{kt}{2}\right)}{\cos^2\left(\frac{kt}{2}\right)+\left(\frac{\omega_0}{k}\right)^2\sin^2\left(\frac{kt}{2}\right)}\right]\left[\cos^2\left(\frac{kt}{2}\right)+\sin^2\left(\frac{kt}{2}\right)\left(\left(\frac{\omega_0}{k}\right)^2-\left(\frac{\gamma}{k}\right)^2\right)+1\right] \text{(A-19)}$$

$$a_4(t) = \frac{1}{2}\left[ \frac{\frac{\gamma^2}{k^2}\sin^2\left(\frac{kt}{2}\right)}{\cos^2\left(\frac{kt}{2}\right)+\left(\frac{\omega_0}{k}\right)^2\sin^2\left(\frac{kt}{2}\right)}\left[\cos^2\left(\frac{kt}{2}\right)+\sin^2\left(\frac{kt}{2}\right)\left(\left(\frac{\omega_0}{k}\right)^2-\left(\frac{\gamma}{k}\right)^2\right)+\frac{\frac{k^2}{\gamma^2}}{\sin^2\left(\frac{kt}{2}\right)}-2\right]+ \right.$$
$$\left. \cos^2\left(\frac{kt}{2}\right)+\sin^2\left(\frac{kt}{2}\right)\left(\left(\frac{\omega_0}{k}\right)^2-\left(\frac{\gamma}{k}\right)^2\right)\right] \qquad \text{(A-20)}$$

where $k = \left(\gamma^2+\omega_0^2\right)^{1/2}$.

**Appendix B**

According to (16) the general form for the decoherecne factor in the Spin-Boson model is:

$$\tilde{D} = \frac{1}{4} \int_0^\infty a_1(t)dt \int_0^\infty \omega_e e^{\frac{-\omega_e}{\lambda}} \cos\omega_e t \coth\left(\frac{\hbar\omega_e}{2k_B T}\right) \qquad (B-1)$$

where $a_1(t)$ defined in (A-17). The above integral has not an exact analytical solution in its original form. However, the evolution of a quantum particle under the influence of an external magnetic field is a kind of time dependent perturbation problems. So, one can investigate the solution of the (B-1) for the various orders of magnitudes of the coupling constant $\gamma$. Since $\frac{\gamma}{\omega_0} \ll 1$, it follows that:

$$(\frac{\omega_0}{k})^2 = \frac{1}{1+\left(\frac{\gamma}{\omega_0}\right)^2} = 1 - \frac{\gamma^2}{\omega_0^2} + \frac{\gamma^4}{\omega_0^4} - \frac{\gamma^6}{\omega_0^6} + \dots \qquad (B-2)$$

$$(\frac{\gamma}{k})^2 = \frac{\gamma^2}{\omega_0^2(1+\left(\frac{\gamma}{\omega_0}\right)^2)} = \frac{\gamma^2}{\omega_0^2}(1 - \frac{\gamma^2}{\omega_0^2} + \frac{\gamma^4}{\omega_0^4} - \frac{\gamma^6}{\omega_0^6} + \dots) \qquad (B-3)$$

Neglecting the higher orders of $\gamma^2$ one gets the following relations for $a_1(t)$ and decoherence factor respectively, in the high temperature limit where $coth\left(\frac{\hbar\omega_e}{2k_B T}\right) \approx \frac{2k_B T}{\hbar\omega_e}$:

$$a_1(t) = \frac{\gamma^2}{\omega_0^2} \sin^2\left(\frac{kt}{2}\right) \qquad (B-4)$$

$$\tilde{D} = \frac{\pi k_B T \gamma^2}{8\hbar \omega_0^2} (1 - e^{-\frac{k}{\lambda}}) \qquad \text{(B-5)}$$

For a broader range of temperature, one can replace $coth\left(\dfrac{\hbar\omega_e}{2k_B T}\right)$ with its corresponding Teylor's series:

$$coth\left(\frac{\hbar\omega_e}{2k_B T}\right) = \frac{2k_B T}{\hbar\omega_e}\left(1 + \frac{2^2 B_1}{2!}\left(\frac{\hbar\omega_e}{2k_B T}\right)^2 - \frac{2^4 B_3}{4!}\left(\frac{\hbar\omega_e}{2k_B T}\right)^4 + \frac{2^6 B_5}{2!}\left(\frac{\hbar\omega_e}{2k_B T}\right)^6 - ...\right) \qquad \text{(B-6)}$$

where $B_1, B_3, ...$ are Bernoulli numbers.

Then the decoherence factor will have the following form:

$$\tilde{D} = \frac{\pi k_B T \gamma^2}{8\hbar \omega_0^2}\left(1 - e^{\frac{-k}{\lambda}} G(T)\right) \qquad \text{(B-7)}$$

where:

$$G(T) = 1 + \frac{2^2 B_1}{2!}\left(\frac{\hbar k}{2k_B T}\right)^2 - \frac{2^4 B_3}{4!}\left(\frac{\hbar k}{2k_B T}\right)^4 + \frac{2^6 B_5}{6!}\left(\frac{\hbar k}{2k_B T}\right)^6 - ... \qquad \text{(B-8)}$$

For the higher orders of $\gamma^2$, $a_1(t)$ could be represented as:

$$a_1(t) = \left[\left(\frac{\gamma^2}{\omega_0^2} - \frac{\gamma^4}{\omega_0^4} + \frac{\gamma^6}{\omega_0^6} - ...\right)\right]\sin^2\frac{kt}{2} - \frac{\left(\frac{\gamma^4}{\omega_0^4} - \frac{\gamma^6}{\omega_0^6} + ...\right)\sin^4\frac{kt}{2}}{1 - \left(\frac{\gamma^2}{\omega_0^2} - \frac{\gamma^4}{\omega_0^4} + \frac{\gamma^6}{\omega_0^6} - ...\right)\sin^2\frac{kt}{2}} \qquad \text{(B-9)}$$

Since, $\left(\dfrac{\gamma^2}{\omega_0^2} - \dfrac{\gamma^4}{\omega_0^4} + \dfrac{\gamma^6}{\omega_0^6} - ...\right)\sin^2\dfrac{kt}{2} \ll 1$, it follows that:

$$\frac{1}{1-\left(\dfrac{\gamma^2}{\omega_0^2}-\dfrac{\gamma^4}{\omega_0^4}+\dfrac{\gamma^6}{\omega_0^6}-...\right)\sin^2\dfrac{kt}{2}}=1+\left(\dfrac{\gamma^2}{\omega_0^2}-\dfrac{\gamma^4}{\omega_0^4}+\dfrac{\gamma^6}{\omega_0^6}-...\right)^2\sin^4\dfrac{kt}{2}$$

$$+\left(\dfrac{\gamma^2}{\omega_0^2}-\dfrac{\gamma^4}{\omega_0^4}+\dfrac{\gamma^6}{\omega_0^6}-...\right)^4\sin^8\dfrac{kt}{2}+...$$

(B-10)

Then, it is legitimate to replace the second term in the right hand side of (B-9) with:

$$\frac{\left(\dfrac{\gamma^4}{\omega_0^4}-\dfrac{\gamma^6}{\omega_0^6}+...\right)\sin^4\dfrac{kt}{2}}{1-\left(\dfrac{\gamma^2}{\omega_0^2}-\dfrac{\gamma^4}{\omega_0^4}+\dfrac{\gamma^6}{\omega_0^6}-...\right)\sin^2\dfrac{kt}{2}}\approx\left(\dfrac{\gamma^4}{\omega_0^4}-\dfrac{\gamma^6}{\omega_0^6}+...\right)\sin^4\dfrac{kt}{2}$$

(B-11)

Using the above approximations one gets the below relation for the decoherence factor:

$$\tilde{D}=\frac{\pi k_B T}{8\hbar}\left[\left(\frac{\gamma^2}{\omega_0^2}-\frac{\gamma^4}{\omega_0^4}+\frac{\gamma^6}{\omega_0^6}-...\right)\left(1-e^{\frac{-k}{\lambda}}G\left(T\right)\right)+\right.$$

$$\left.\left(\frac{\gamma^4}{\omega_0^4}-\frac{\gamma^6}{\omega_0^6}+...\right)\left(\frac{1}{4}\cosh\left(\frac{2k}{\lambda}\right)\overset{\cdot}{G}\left(T\right)-\cosh\left(\frac{k}{\lambda}\right)G(T)+\frac{3}{4}\right)\right]$$

(B-12)

where:

$$\overset{\cdot}{G}\left(T\right)=1+\frac{2^4 B_1}{2!}\left(\frac{\hbar k}{2k_B T}\right)^2-\frac{2^8 B_3}{4!}\left(\frac{\hbar k}{2k_B T}\right)^4+\frac{2^{10}B_5}{6!}\left(\frac{\hbar k}{2k_B T}\right)^6-...$$

(B-13)